\documentclass{JHEP3}
\usepackage{graphicx}
\usepackage{amsmath}
\usepackage{amsfonts}
\usepackage{amssymb}
\newcommand{\be}{\begin{equation}}
\newcommand{\ee}{\end{equation}}
\newcommand{\ben}{\begin{eqnarray}}
\newcommand{\een}{\end{eqnarray}}

\title{Locally Localized Gravity and Geometric Transitions}

\author{Dionisio Bazeia,$^a$ Francisco A. Brito,$^b$ and Adalto Rodrigues Gomes$^{ac}$
\\
$^a$Departamento de F\'\i sica, Universidade Federal
da Para\'\i ba,\\
Caixa Postal 5008, 58051-970 Jo\~ao Pessoa, Para\'\i ba, Brazil
\\
$^b$Departamento de F\'\i sica, Universidade Federal de Campina Grande,
\\
58109-970 Campina Grande, Para\'\i ba, Brazil
\\
$^c$Departamento de Ci\^encias Exatas, Centro Federal de
Educa\c c\~ao Tecnol\'ogica do Maranh\~ao,
65025-001 S\~ao Lu\'\i s, Maranh\~ao, Brazil\\
E-mails: bazeia@fisica.ufpb.br, fabrito@df.ufcg.edu.br, argomes@fisica.ufpb.br}

\abstract{{In this paper we analyze the local localization of gravity in $AdS_4$
thick brane embedded in $AdS_5$ space. The 3-brane is modelled by domain wall
solution of a theory with a bulk scalar field coupled to five-dimensional gravity.
In addition to small four-dimensional cosmological constant, the vacuum expectation
value (vev) of the scalar field controls the emergence of a localized four-dimensional
quasi-zero mode. We introduce high temperature effects, and we show that gravity
localization on a thick 3-brane is favored below a critical temperature $T_c$.
These investigations suggest the appearance of another critical temperature $T_*,$
where the thick 3-brane engenders the geometric $AdS/M/dS$ transitions.\\

\vspace{1cm}

Keywords: Field Theories in Higher Dimensions, Classical Theories of Gravity.}}

\maketitle

\begin{document}

\section{Introduction}

The localization of gravity on a brane \cite{rs} has appeared as
an alternative to compactification involving infinite extra
dimension. This realization can be considered in a five-dimensional
gravity theory, with a negative cosmological constant $\Lambda_5,$ where
we can obtain a 5d anti-de-Sitter ($AdS_5$) solution. When a 3-brane is introduced,
gravity can be localized on it. In the Randall-Sundrum \cite{rs,rs1} scenarios,
3-branes are embedded in $AdS_5$ bulk space where one considers a five-dimensional
gravity with negative cosmological constant $\Lambda_5$ and source of ``infinitely thin''
3-branes given by delta functions. As it was shown, there is a {\it perfect} fine-tuning
between branes tension and the cosmological constant $\Lambda_5$. This fine-tuning leads
to a 4d Minkowski ($M_4$) brane with four-dimensional cosmological constant ${\Lambda}=0$,
such that only the $AdS_5$ space is curved. The graviton zero mode
bound to the 3-brane is responsible for a 4d localized gravity.
The correction to the Newtonian potential due to Kaluza-Klein
gravitons is highly suppressed, at low energy. On the other hand, if perfect
fine-tunings are absent then both 3-brane and $AdS_5$ space can be
curved. These branes are either 4d de Sitter ($dS_4$) branes with
${\Lambda}>0$ or 4d anti-de-Sitter ($AdS_4$) branes with
${\Lambda}<0$. Explicit solutions of $AdS_4$, $dS_4$ and $M_4$ branes were put
forward in \cite{mc93,mc99,kaloper,kin,nihei,wfgk,kr}.

The issue of local localization of gravity on $AdS_4$ branes was
firstly addressed in \cite{kr} --- see also
\cite{porrati,kogan,kogan2} for connection between massive gravity
in $dS_4$ and $AdS_4$ space and absence of van
Dam-Veltman-Zakharov discontinuity. The graviton mode responsible
for the 4d gravity is not a zero mode but an almost massless mode,
the ``quasi-zero'' mode that dominates over the
Kaluza-Klein modes. This is a far more general mechanism of
localization of gravity, because it does not require any condition on
the space far from the brane. Under this perspective, gravity
localization can be realized in string theory, once no-go theorems
about localization in supergravity theories, e.g.
\cite{linde,cvetic,malda} relying on the asymptotic behavior of
the geometry do not necessarily apply \cite{kr}--- see \cite{kr2}
for further discussions. There are partial supersymmetric realizations of massless
\cite{clp} and massive \cite{justin} gravity localization in braneworld scenarios,
which arise from a sphere reduction in string/M-theory. Here, only the bulk
Lagrangian can be viewed as arising from a higher dimensional supergravity while
the 3-brane is supported by a delta function put by hand. However, as one considers
non-homogeneous quaternionic moduli spaces, a complete 5d supergravity realization
of gravity localization can be incorporated smoothly by thick 3-branes supported by
scalar fields \cite{kbd}. Thus, under this perspective it is useful to study models
where thick brane solutions exist, and gravity can also be trapped.

Universal aspects of localization of gravity in {\it thick} $M_4$ branes were first
studied in \cite{csaki}. Scalar fields to model such thick branes were introduced
in \cite{wfgk,csaki,gremm}. Thick $dS_4$ and $AdS_4$ branes which use bulk
scalar fields have also been considered in the literature. Models implementing
a scalar field with scalar potentials like $\cos^{n}(\phi)$ that can
be solved analytically have been investigated in
\cite{ira_shiu,koba_soda,wang,Naoki,Naoki2,oscar} in different
contexts. A $\lambda\phi^4$ model was explored analytically in
\cite{wfgk} in first order formalism, but just in the
${\Lambda}=0$ case. In Ref.~\cite{ioda} this model was also
considered in the context of local localization of gravity, but
the solutions were found only in the thin wall limit.

In this paper we consider a $\lambda\phi^4$ model to investigate
the local localization of gravity on {\it thick} $AdS_4$ branes.
We use numerical methods to solve the equations of motion and to
obtain the graviton spectrum of gravity fluctuations around the
brane solution. We explore in detail how the lightest graviton
mode binds to the brane. The scalar field vev and the coupling constant
$\lambda$ control the emergence of this graviton. We also consider high
temperature effects in the bulk and discuss how they can affect the localization
of gravity. Our motivation is to address the issue of geometric transitions, where
the 3-brane changes from $AdS_4$ to $M_4$, and then to $dS_4$ as the temperature diminishes.
This mechanism occurs on a thick 3-brane, lifting a supersymmetric vacuum ${\Lambda}<0$
to another one, non-supersymmetric, with ${\Lambda}>0.$ This is a current discussion
in string theory; see, e.g., the investigation introduced in Ref~{\cite{kklt}}. A small and
positive 4d cosmological constant agrees with current observational data, which show that our
universe experiences an accelerated expansion \cite{ac,ac2}.

The paper is organized as follows. In Sec.~\ref{prml} we introduce
the model and the formalism applied to our analysis. In
Sec.~\ref{thin} we discuss how brane solutions appear in a theory
engendering 5d gravity and scalar field in the thin wall limit. In
Sec.~\ref{thick} we extend the analysis of the previous section to
the study of thick branes and high temperature effects.  We end the paper
in Sec.~\ref{conclu} where we present our final considerations.

\section{Preliminaries}
\label{prml}

We consider the model with five-dimensional gravity coupled to a
scalar field \ben \label{action} S=\int{d^{\,5}x\sqrt{g}
\left[-\frac{1}{4}R +
\frac{1}{2}\partial_M\phi\partial^M\phi-V(\phi)\right]},\een where
we consider the signature (+ - - - -) and $M\!=\!0,1,2,3,4,$ with $g=\det(g_{MN})$.

The metric Ansatz is \ben
\label{metric}  ds^2=e^{2A(r)}\bar{g}_{\mu\nu}dx^\mu dx^\nu -
dr^2, \een where $\bar{g}_{\mu\nu}$ is the four-dimensional
metric, with $\mu,\nu=0,1,2,3$, satisfying
\ben\label{Rbar}\bar{R}_{\mu\nu}=-3{\Lambda}\bar{g}_{\mu\nu}.\een
The four-dimensional cosmological constant ${\Lambda}$ is positive for de
Sitter ($dS_4$) spacetime, negative for anti-de Sitter ($AdS_4$)
spacetime and zero for Minkowski ($M_4$) spacetime.

We are mainly interested in localization of gravity on a 3-brane. This requires
the study of gravity fluctuations around the brane solution \cite{rs,wfgk,kr,csaki}.
On this account we linearize the Einstein equations by considering a perturbation
as $\bar{g}_{\mu\nu}=g_{\mu\nu}+h_{\mu\nu}$. Taking into account only
the traceless transverse (TT) sector of the linear perturbation,
where gravity equations of motion do not couple to matter fields,
we find $\partial_M(\sqrt{g}g^{MN}\partial_N)\Phi=0$ or in terms
of metric components \ben\label{Phi}
\left(\partial_r^2+4A'\partial_r-e^{-2A}(\Box_{4d}+2{\Lambda})\right)\Phi=0,
\een where $\Phi$ describes the wave function of the graviton on
non-compact coordinates. Let us consider the Ansatz
$\Phi=h(r)M(x^\mu)$,  the equation describing the 4d graviton
$(\Box_{4d}+2{\Lambda})M\!=\!m^2M$ and the change of variable
$h(r)=e^{3A(z)/2}\psi(z)$, $z(r)=\int{e^{-A(r)}dr}$ into
eq.~(\ref{Phi}). In this way we get to the Schroedinger-like
equation
\ben\label{sch}-\partial_z^2\psi(z)+{U}(z)\psi(z)=m^2\psi(z),
\een
with the potential
\ben
\label{Vz}{U}(z)=\frac{9}{4}A'(z)^2+\frac{3}{2}A''(z).
\een

\section{Brane solutions in the thin wall limit}
\label{thin}

To show how a $Z_2$-symmetric 3-brane arises
through the scalar field in the thin wall limit \cite{wfgk,bcy},
we consider the equations of motion
\ben \label{eom1}
\phi''+4A'\phi'&=&\frac{\partial V(\phi)}{\partial\phi}\\
\label{eom2} A''+{\Lambda}e^{-2A}&=&-\frac{2}{3}\phi'^2\\
\label{eom3}{A'}^2-{\Lambda}e^{-2A}&=&
-\frac{1}{3}V(\phi)+\frac{1}{6}{\phi'}^2
\een
where we assume that the scalar field only depends on the extra dimension, $r$.
For non-zero cosmological constant ${\Lambda},$ the integrability of these equations
through first order formalism \cite{wfgk} is hard. For this reason, we work with
the second order equations.

In our analysis, the scalar field is a typical kink approaching the
asymptotic values $\phi(r\to\pm\infty)\to\pm a$. We choose the kink
solution and the $Z_2$-symmetric scalar potential as
\ben
\label{thin_kink} \phi&=&a\tanh{\lambda ar}\\
\label{scal_pot}
V(\phi)&=&\frac12{\lambda^2}(\phi^2-a^2)^2-\frac{3}{L^2},
\een
where $V(\pm a)\equiv \Lambda_5=-3/L^2$ is identified with
the $AdS_5$ cosmological constant. We can quickly check that this
cannot even satisfy the equation (\ref{eom1}) because of the term
$4A'\phi'$. However, this is not true in the
{\it thin wall limit}. In this limit ($\lambda\to\infty$ and $a\to0,$
with $\lambda a^3$ fixed) we can approach the kink solution to a
step function $\phi\simeq a\,{\rm sgn}(r)$ whose width
$\Delta\simeq 1/\lambda a$ goes to zero. This provides the
``identities''
\ben \label{deltas}\phi'\simeq 2a\delta(r), \qquad
{\phi'}^2\simeq\sigma\delta(r).
\een
In our five-dimensional set up, $\sigma$ is identified with a positive
3-brane tension given by
\ben\label{tension}
\sigma=\frac{4}{3}\lambda a^3.
\een
This is precisely the kink energy in absence of gravity.
As we shall see
later, we can also use this formula for thick branes coupled to 5d gravity.
This is because the back reaction on the kink by turning on gravity is
very small.

We now turn our attention to equations
(\ref{eom1})-(\ref{eom3}). In the thin wall limit
$A'\phi'\to2 a A'(r)\delta(r)$, which is zero everywhere provided the
function $A(r)$ satisfies the boundary condition $A'(0)=0$. In
this limit, the equation (\ref{eom1}) is satisfied and the two
other equations now reads
\ben
\label{eom1.2} A''+{\Lambda}e^{-2A}&=&-\frac{2}{3}\sigma\delta(r)\\
\label{eom2.2} {A'}^2-{\Lambda}e^{-2A}&=&\frac{1}{L^2}.
\een
Note that the scalar field contributions in (\ref{eom3}) cancel out, because of their
consistency with (\ref{eom1}). The equations above could be
obtained from the action
\ben \label{thin_action}
S=\int{d^{\,5}x\sqrt{g} \left[-\frac{1}{4}R +
\frac{3}{L^2}\right]}-\sigma\int{d^{\,4}x
dr\sqrt{g}\,\delta(r)},
\een
which is the thin wall limit of the
action (\ref{action}). It describes two copies of $AdS_5$ with
curvature $\Lambda_5=-3/L^2$ pasted together along an ``infinitely
thin'' 3-brane located at $r=0$ with tension $\sigma$.

It is not difficult to check that the equations
(\ref{eom1.2})-(\ref{eom2.2}) are satisfied by the well known
solutions \cite{mc93,mc99,kaloper,kin,nihei,wfgk,kr}
\ben\label{dS_4}
&&\mbox{$dS_4$ (${\Lambda}>0)$:}\qquad
A(r)=\ln\left(\sqrt{{\Lambda}}L\sinh\frac{c-|r|}{L}\right),\qquad
\sigma=\frac{3}{L}\coth{\frac{c}{L}} \\
\label{M_4} &&\mbox{$M_4$
(${\Lambda}=0)$:}\qquad A(r)=\frac{c-|r|}{L},\qquad \sigma=\frac{3}{L}\\
\label{AdS_4} &&\mbox{$AdS_4$ (${\Lambda}<0)$:}\qquad
A(r)=\ln\left(\sqrt{-{\Lambda}}L\cosh\frac{c-|r|}{L}\right),\qquad
\sigma=\frac{3}{L}\tanh{\frac{c}{L}},
\een
where $c$ is a real constant. As we mentioned earlier,
the perfect fine-tuning $\sigma=3/L=L|\Lambda_5|$ in the $M_4$
brane imposes ${\Lambda}=0$. By making $c\to\infty$ in the
other fine-tuning equations,  both $dS_4$ and $AdS_4$ branes
collapse to $M_4$ branes. This is precisely the fine-tuning
imposed in the Randall-Sundrum scenario \cite{rs1}. The 3-brane
tension is now given explicitly in terms of the vev $a$ of the
scalar field via equation (\ref{tension}). There exists an analog
of this picture in four-dimensional supergravity domain walls.
The $M_4$ brane with perfect
fine-tuning, i.e., $c\to\infty$, is analogous to a 4d BPS
saturated domain wall ($\sigma=\sigma_{BPS}$) while the $dS_4$ and
$AdS_4$ branes with $c$ finite are analogous to 4d ``bent walls''
whose tension are $\sigma>\sigma_{BPS}$ and $\sigma<\sigma_{BPS}$,
respectively --- for a review see \cite{mirjam_soleng,mc2000}.

The physics of such solutions as gravity localization on $AdS_4$
branes was first explored in \cite{kr} --- see also
\cite{miemiec,mat_schwartz}. As it was stressed, this is the scenario
where {\it locally localized gravity} emerges. This phenomenon is local,
because in the solution (\ref{AdS_4}), it is only around $|r|\leq c$ that
the warp factor approaches that of a $M_4$ brane. For $|r|\gg c$ the space
include the $AdS$ boundary, so it has infinite volume. By
showing that gravity localization is a local effect, we avoid
issues concerning the global behavior of the extra dimension.
Furthermore, since the volume of extra dimension is infinite in
this case, the normalized graviton bound to the brane cannot be a
massless one. In fact, as it has been shown, the graviton bound to
the brane is an almost massless graviton \cite{kr} --- for phenomenological
issues on extra dimensions with infinite volume see
Refs.~{\cite{gregory,dvali1,dvali2}}.

In the following sections, we search for thick brane solutions and
explore how this physical scenario emerges in our model.

\section{Thick brane solutions}
\label{thick}

In this section, we relax the limit of zero thickness considered
above to find how thick branes can arise via the scalar field. We
shall essentially look for a smooth version of the ``infinitely thin'' brane
solutions (\ref{dS_4}), (\ref{M_4}) and (\ref{AdS_4}) and study
how they can localize gravity. We shall mainly focus on gravity
localization in $AdS_4$ branes. For thick branes, however, due to
difficulties to integrate eqs.~(\ref{eom1})-(\ref{eom3}) analytically,
we content ourselves with numerical solutions.

We deal mainly with eqs.~(\ref{eom1})-(\ref{eom2}), while eq.~(\ref{eom3}), as an
energy conservation check, give us $\Lambda_5.$ The boundary conditions
$\phi(0)=0,\phi(\infty)=a$ and $A(0)=0, A'(0)=0$ suggest that we use the finite difference
method for $\phi(r),$ and Runge-Kutta method  for $A(r)$. We propose the kink profile
$\phi_1(r)=a\tanh(\lambda ar)$ as a first tentative solution. Then we use $\phi_1(r)$ in
eq.~(\ref{eom2}) and apply Runge-Kutta method to obtain $A_1(r).$ Next, we transform
eq.~(\ref{eom1}) in a set of $N$ coupled nonlinear algebraic equations, with $A_1(r)$
given, and with $N$
values $\phi_2(r)$ to be determined. The procedure goes on until a desired precision
to obtain $\phi(r)$ and $A(r)$ is achieved \cite{num1,num2,num3}. Then, we use
eq.~(\ref{eom3}) to get $\Lambda_5.$ As an example, for $\lambda=a=1$ and $\Lambda=-0.2,$
as we go further with the iterations, eq.~(\ref{eom3}) leads $\Lambda_5$
to the value $-0.77.$

\begin{figure}[ht]
\begin{center}
\includegraphics[{angle=90,height=8cm,angle=180,width=9cm}]{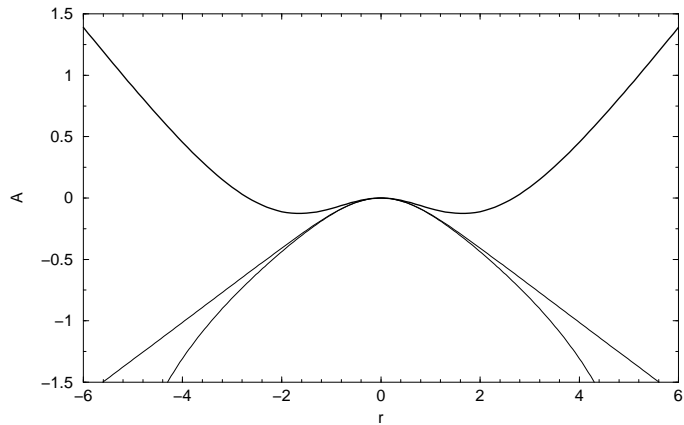}
\end{center}
\caption{The solution $A(r)$ for ${\Lambda}=-0.2$ (thicker curve),
${\Lambda}=0$, and ${\Lambda}=+0.02$ (thinner curve).}\label{fig0.1}
\end{figure}

\begin{figure}[ht]
\begin{center}
\includegraphics[{angle=90,height=8.0cm,angle=180,width=10cm}]{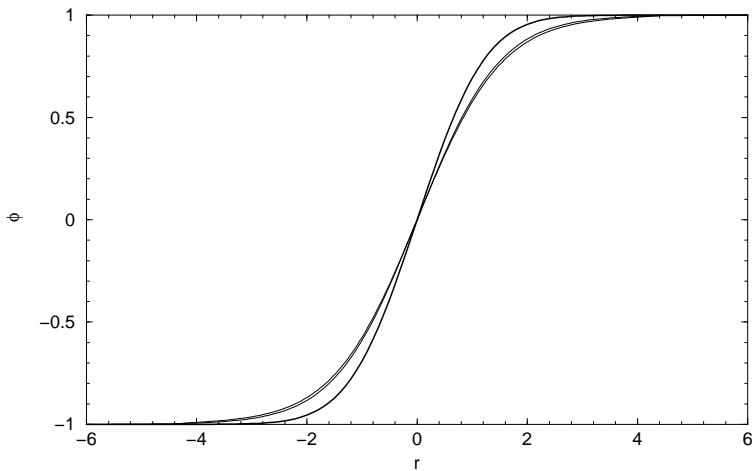}
\end{center}
\caption{The kink solutions $\phi(r)$ for ${\Lambda}=-0.2,\,0,$ and $+0.02$.
They almost confuse in the last two cases.}\label{fig0.2}
\end{figure}

\begin{figure}[ht]
\begin{center}
\includegraphics[{angle=90,height=8.0cm,angle=180,width=9.0cm}]{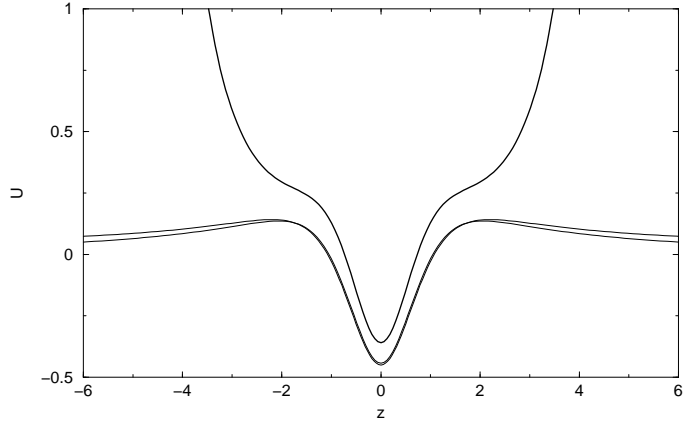}
\end{center}
\caption{The Schroedinger-like potential for
${\Lambda}=-0.2,\,0,$ and $+0.02$.}\label{fig0.3}
\end{figure}

The parameters $a$ and $\lambda$ are specified according to the thickness and
tension of the brane. The cosmological constant ${\Lambda}$ is
also specified with the scenario we want to study, i.e., $AdS_4$,
$dS_4$ or $M_4$. In Fig.~\ref{fig0.1} and in Fig.~\ref{fig0.2} we
show the behavior of the solutions $A(r)$ and the kink profile
$\phi(r)$, respectively, for ${\Lambda}=-0.2,\,0,\,+0.02$.
Note that the kink profiles are almost the same regardless of the value
of ${\Lambda}$, with slightly smaller thickness in the $AdS_4$ case.
The tension can be related to ${\Lambda}$ and to $\Lambda_5$ via fine-tuning
\cite{rs,rs1,kr,mirjam_soleng,mc2000}. Indeed, there exists a function that can
be determined numerically engendering a fine-tuning
$\sigma=(3/L)f(c)$. For $\lambda\!=\!a\!=\!1$ we find $f(c)=0.87$,
where $c\simeq 1.6$ for ${\Lambda}=-0.2$. A finite constant
$c$ characterizes the solution $A(r)$ for non-zero 4d cosmological
constant --- see the solutions in eqs.~(\ref{dS_4}) and (\ref{AdS_4}). As also happens
in infinitely thin branes \cite{kr}, for $dS_4$ branes $c$ is the distance between
the brane and the horizon, whereas for $AdS_4$ branes $c$ is the distance to the
turn around point in $A(r)$ --- see Fig.~\ref{fig0.1}. The latter
case is of particular interest. The solution diverges for $r\gg c$
approaching the boundary of the $AdS_5,$ but close to the brane
($r\lesssim c$) it behaves like a $M_4$ brane. Thus, it is expected
that gravity is locally localized on the 3-brane. The bound
graviton, however, is not a massless graviton, but an almost
massless graviton. In fact, this graviton is the lightest mode of
an infinite Kaluza-Klein tower of states as the Schroedinger-like
potential for ${\Lambda}=-0.2$ indicates in Fig.~\ref{fig0.3}.
This potential essentially represents a box in contrast to the
volcano-like potentials for ${\Lambda}=0$ and
${\Lambda}=+0.02$. In the $dS_4$ case, the potential asymptotes
to non-zero value, showing that there is a gap separating the
zero mode from the continuum spectrum. In the following we turn our
attention only to the $AdS_4$ case.

\subsection{The localization of gravity in $AdS_4$ thick branes}

The results show that as we decrease the brane tension keeping the
brane thickness fixed, in the limit of tensionless brane we reproduce the spectrum
of gravity fluctuations of pure $AdS_5$ space \cite{kr,miemiec,mat_schwartz}. The numerical
calculations below are performed using ${\Lambda}=-0.2$, except in some cases where we have
to specify other values. 

\begin{figure}[ht]
\begin{center}
\includegraphics[{angle=90,height=8.0cm,angle=180,width=9.0cm}]{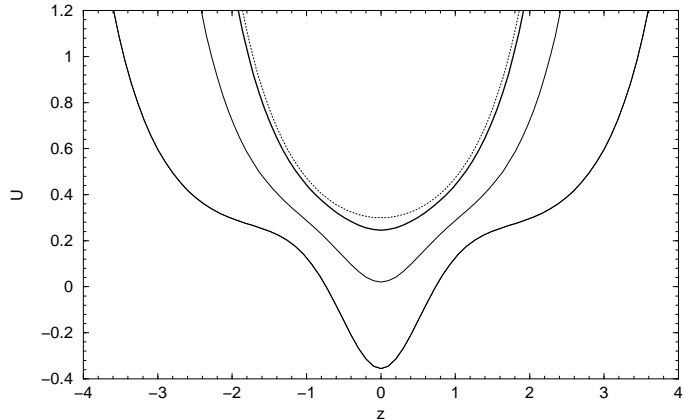}
\end{center}
\caption{The Schroedinger-like potential for fixed width and
tensions decreasing with $a$=1, 5/10 and 2/10. For $a=2/10$ we
approach the tensionless brane limit (upper curve).}\label{fig1.1}
\end{figure}

In Fig.~\ref{fig1.1} we show how the Schroedinger-like potential
changes with the brane tension. In the tensionless brane limit we have a pure
$AdS_5$ potential. In this limit, the solution of the quantum mechanics problem
gives us the energy eigenvalues \cite{kr}
\ben
\label{spct5} E=n(n+3), \qquad n=1,3,5...
\een
The squared masses are defined as $m^2=E|{\Lambda}|.$ The other cases are harder to
investigate. However, we can find the energy spectrum numerically by searching for the
zeroes of the wavefunction $\psi$ at the boundaries of the box-like potential $U(z)$
as a function of the energy $E$, as suggested in Ref.~\cite{kr}.
For each energy, the wavefunction is obtained by the Numerov method \cite{mueller}.

In our investigation, we have found an interesting behavior: as we increase the
brane tension, one mode become almost massless in comparison with all the other
modes of the spectrum. It has an amplitude on the brane much higher than
all the other modes, as we shall see later explicitly. We refer to this
state as the {\it quasi-zero} mode. This is the mode responsible for 
localizing gravity on the brane.

\begin{figure}[ht]
\begin{center}
\includegraphics[{angle=90,height=8.0cm,angle=180,width=9.0cm}]{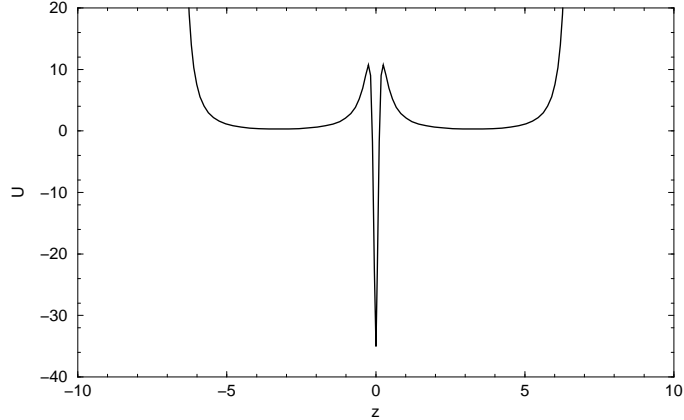}
\end{center}
\caption{The Schroedinger-like potential
for a brane with $\Delta\!=\!0.1$ and tension for $a=1.0$. The
$AdS_5$ potential tends to split into two copies.}\label{fig1.2}
\end{figure}

In the thin wall approximation $(\Delta=0.1,\,{\rm for\; instance})$, as we
increase the tension, the Schroedinger-like potential tends to become two copies
of the pure $AdS_5$ potential pasted together along the brane source --- see
Fig.~\ref{fig1.2}. This is similar to the critical limit of brane tension found
in \cite{kr,miemiec,mat_schwartz}. The physical meaning is that now in the $AdS_5$
spectrum there is one quasi-zero mode trapped on the brane responding for the
four-dimensional gravity.

\begin{figure}[ht]
\begin{center}
\includegraphics[{angle=90,height=7.0cm,angle=180,width=8.0cm}]{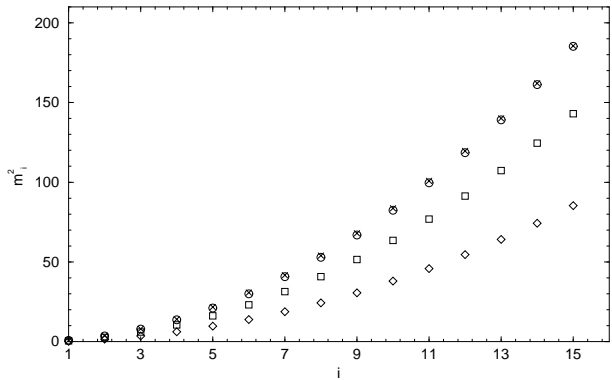}
\end{center}
\caption{Graviton masses for a fixed brane thickness $\Delta=1$
 and tensions $\sigma=(4/3)a^2$ for $a$=1 ($ diamonds$), 5/10 ($squares$),
 2/10 ($circles$) and $a=0$ ($crosses$).}\label{fig6}
\end{figure}

\begin{figure}[ht]
\begin{center}
\includegraphics[{angle=90,height=7.0cm,angle=180,width=8.0cm}]{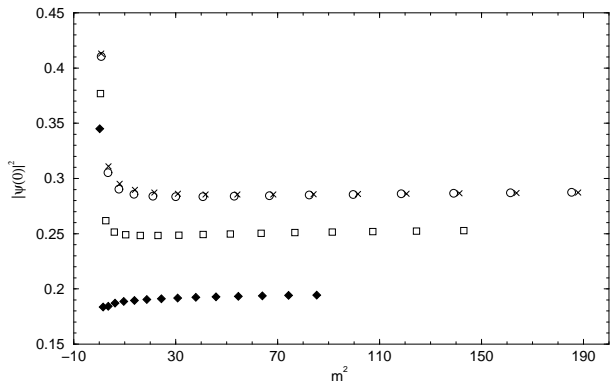}
\end{center}
\caption{The wavefunction amplitude on the brane $|\psi_i(z=0)|^2$
for the squared masses $m_i^2,$ with brane thickness $\Delta=1$ and tensions
$\sigma=(4/3)a^2.$ The plots are for $a$=1 ($black\,diamonds$), 5/10 ($squares$), 2/10
($circles$), and $a=0$ ($crosses$).}\label{fig7}
\end{figure}

Note that by fixing $\Delta=1/\lambda a,$ the tension
$\sigma=(4/3)a^2/\Delta$ depends only on $a$. In this way, we can
leave only the scalar field vev $a$ as a free parameter. This
parameter can be understood as an energy scale that can control
the localization of gravity. As an example, for $\Delta=1$ and
several brane tensions $\sigma\!=\!(4/3)a^2$, Fig.~\ref{fig6} shows
how in the tensionless brane limit the massive graviton spectrum
obtained approaches the pure $AdS_5$ spectrum just as in the
Karch-Randall scenario (crosses). For numerical reference see the
Table IV.A. For higher values of brane tensions, the
wavefunction amplitude of the first mode is very much larger than
the amplitude of the other excited states. In Fig.~\ref{fig7} we show that
the lightest mode is at $a=1,$ since it occurs for greater tension; the first mode
(upper black diamond) has amplitude on the brane much higher than the other modes
(adjacent black diamonds). It separates from the other modes, and it is our almost massless
mode that contribute to the Newtonian potential as a leading term that dominates over
the other Kaluza-Klein (heavier) modes. Another behavior shown in Fig.~{\ref{fig7}} is that
as the squared masses $m_i^2$ increase, the amplitudes $|\psi_i(z=0)|^2$ tend to constants
which depend of the corresponding tensions. In fact, in this limit we have checked that
the wavefunctions for the potential $U(z)$ can be very well approximated to that of a
box-like potential, where $|\psi(0)|^2=1/z_{\infty},$ with $z_{\infty}$ being a constant
that identifies the contour of the box.

\begin{figure}[ht]
\begin{center}
\includegraphics[{angle=90,height=7.0cm,angle=180,width=8.0cm}]{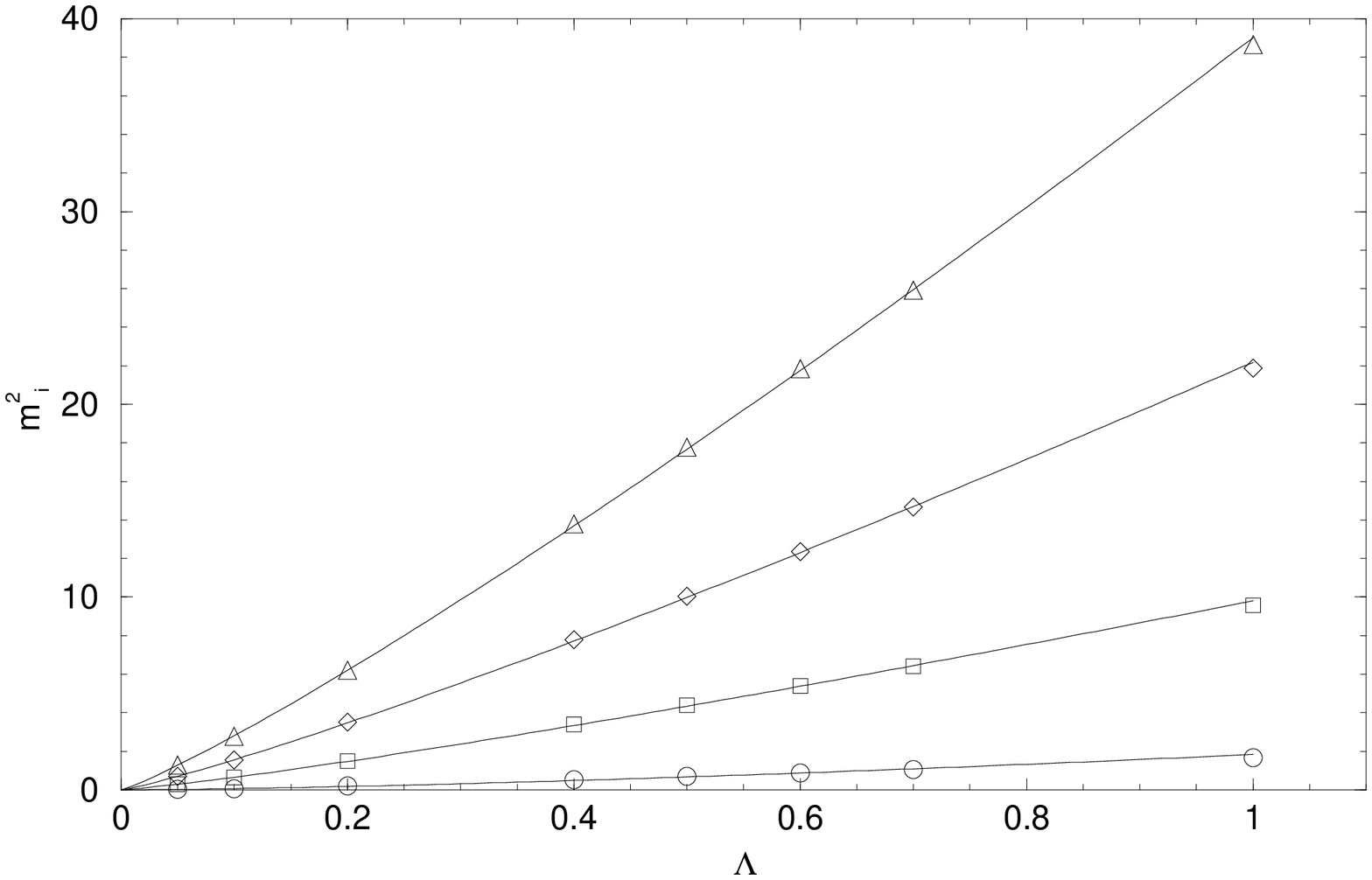}
\end{center}
\caption{The square of mass of the four first graviton modes as a function of
${\Lambda}$ for a fixed brane thickness $\Delta=1$ and brane tension
$\sigma=(4/3)$.}\label{mLambda}
\end{figure}

We can also see that as $|{\Lambda}|$ decreases, the mass squared of 
this graviton mode, the quasi-zero mode $m_1^2$ vanishes faster than all the other modes,
as we show in Fig.~\ref{mLambda}. To illustrate this situation, we recall that the mass
squared of the first four modes depends on $|{\Lambda}|$ according to the
power-laws:
\ben
m_1^2\simeq 1.83\,|{\Lambda}|^{1.46}, \:\:\: m_2^2\simeq9.80\,|{\Lambda}|^{1.18},
\:\:\: m_3^2\simeq22.20\,|{\Lambda}|^{1.15},\:\:\: m_4^2\simeq39.0\,|{\Lambda}|^{1.14},
\een
See also \cite{miemiec,mat_schwartz} for former investigations on this issue. We have
checked that although the power in $|\Lambda|$ decreases for increasing $i,$
it remains always bigger than $1.$ This is in agreement with the fact that in the limit
${\Lambda}\to0,$ we approach the $M_4$ case, in which the brane is flat and the graviton
is a zero mode \cite{rs}, with no other discrete mode in the spectrum. The massless
limit is smooth, and shows no van Dam-Veltman-Zakharov
discontinuity \cite{porrati,kogan,kogan2}. Thus, the brane
tension and 4d cosmological constant ${\Lambda}$ are responsible
for the emergence of the almost massless graviton. Similar results
are obtained by fixing the brane tension and varying the brane thickness.

\begin{table}[ht]
\begin{center}
\begin{tabular}
{||c||c|c|c|c||}\hline\hline
$i$ & a=1 & a=5/10 & a=2/10 & $AdS_5$ \\
\hline\hline
1 & 0.190&  0.507&0.753&  0.8 \\
\hline\hline
2&  1.493&  2.618&3.474&  3.6\\
\hline\hline
3&  3.488&  5.917&7.751&  8.0\\
\hline\hline
4&  6.190&  10.417&13.592&  14.0\\
\hline\hline
5&  9.622&  16.133&21.011&  21.6\\
\hline\hline
6&  13.784&  23.082&30.025&  30.8\\
\hline\hline
\end{tabular}
\end{center}
\caption{Graviton masses for decreasing tensions $\sigma=(4/3)a^2$ with
$\Delta=1$, $a=1,5/10,2/10$. The last column is for pure $AdS_5$ masses
$m_i^2=n(n+3)|{\Lambda}|$, $n=2i-1$.}
\end{table}

\subsection{High temperature effects and geometric transitions}

In cosmology, the cosmic evolution may be related to the temperature, which is
introduced as a parameter that controls the way a phase transition can occur. 
In our investigation, the vev scale $a$ that we
have used can perfectly be related to high temperature effects. The 5d fields in the
bulk can be regarded as 4d fields on the brane with an infinite Kaluza-Klein (KK)
tower of masses. On this account one can compute the 5d effective potential at high
temperature in the standard way by summing the contribution of the KK tower to the
4d effective potential \cite{odintsov}.

As we know, the effective mass for a scalar meson at high temperature has the
standard behavior \cite{dja,wei}
\ben\label{temp}
m^2(T)=m^2(0)+B T^2,
\een
where $B$ is a numerical coefficient that depends on the higher excited Kaluza-Klein
modes \cite{odintsov}. In general, there are two cases to be considered: first, $m^2(0)<0$
and $B>0,$ where the scalar potential has its symmetry restored at temperatures higher than
a critical temperature $T_c,$ and, second, $m^2(0)>0$ and $B<0,$ where symmetry is restorated
at temperatures lower than $T_c$. In Ref.~\cite{vilenkin}, it was addressed the issue of
disappearance of 2d domain walls embedded in a 4d Minkowski space-time. This requires that
one assumes the second case, where symmetry restoration occurs at temperatures lower than
$T_c$. In this regime, there is no domain wall at all at low temperature, as it is required
by observational data in our 4d universe. However, if we consider that our universe
is itself a 3d domain wall that was formed at very high temperature, we necessarily
have to follow the route that appears in the first case above.

Furthermore, in the former section we considered a domain wall with a negative
cosmological constant. This 3d domain wall is what we referred to as an $AdS_4$ thick
3-brane. For this reason, our main goal in this subsection is to investigate how the brane
tension, which should depend on the temperature, could be used to control localization of
gravity. Moreover, since the cosmological constant depends on the brane tension, we shall
also study how it can induce a phase transition driven by the 4d cosmological
constant ${\Lambda}$.

The determination of the value of $B$ is model dependent, and is out of the scope of this
paper --- see \cite{odintsov} for further details.\footnote{Indeed, in the region of
criticality, that is, for $T$ around $T_*,$ we have that $\Lambda$ is around zero
and the results obtained in \cite{odintsov} for flat 3-brane can also be applied here.}
In spite of this, we can go on to see that in our model we have $m^2(0)=-2\lambda^2a^2;$
thus, at high temperature one breaks the symmetry as the temperature diminishes
until $m^2(T=T_c)=0,$ where a critical temperature $T_c^2=2\lambda^2a^2/B$
is found, which corresponds to the tensionless brane limit. In this limit,
clearly there is no brane. By lowering the temperature to $T<T_c,$ we reach
the symmetry-broken phase of the scalar potential, and then a brane with tension
$\sigma\sim\lambda\,{a_T}^3$ and scale $a_T=\sqrt{a^2-BT^2/2\lambda^2}$ is recuperated.
In this regime, we can find a temperature interval where the tension is large enough
to favor localization of gravity.

As the temperature decreases even more, the tension becomes even larger and it is
expected that in such regime another phase transition occurs, since the tension can
dominate over the 5d cosmological constant. We can achieve this situation from the
fine-tuning equations of Sec.~\ref{thin}~--- which are valid in the thin wall approximation.
The information can be recast to the form \cite{kr}
\ben\label{ftuning}
{\Lambda}=\frac{1}{L^2}\left(\frac{L^2\sigma^2}{9}-1\right).
\een
This result shows that by fixing $L\Lambda_5$ for all temperatures, we have
(i) ${\Lambda}>0$ for $\sigma>\sigma_*$, (ii) ${\Lambda}=0$ for $\sigma=\sigma_*$
and (iii) ${\Lambda}<0$ for $\sigma<\sigma_*$, where $\sigma_*=L|\Lambda_5|$ is the
critical tension. This give us another critical temperature
\ben
\label{2ndT}
{T_*}^2\simeq{T_c}^2-\frac{2}{B}\left({\lambda^2 L|\Lambda_5|}\right)^{2/3},
\een
which is clearly below the critical temperature of brane formation and gravity
localization $T_c$. The $AdS_4$ brane exists and localizes gravity only in the interval
$T_*<T<T_c$. For $T=T_*$ only the $M_4$ brane is allowed, and for $T<T_*$ it is the $dS_4$
brane which is favored. The reasoning is valid under the high temperature approximation,
so we have to care about extending it into a much lower temperature region.

We notice that the geometric transitions $AdS/M/dS$ at $T=T_*$ imply a breaking of
supersymmetry on the 3-brane, since supersymmetry is not compatible with the $dS_4$
structure of the
spacetime. This scenario seems to conform very naturally with phenomenology, and will be
further investigated elsewhere.  

\section{Discussions}
\label{conclu}

In this work, we have investigated graviton localization on a single $AdS_4$ thick brane,
where a quasi-zero mode is responsible for 4d gravity. The brane tension increases
monotonically with the vev of the bulk scalar field. For values of vev large enough
an almost massless mode emerges. In this way, the vev is a scale of energy which can control
localization of gravity. When we turn on the high temperature effects in the bulk, this scale
becomes temperature dependent and so does the tension. Thus, there is a critical temperature
$T_c$ for which there is no brane --- the tensionless brane limit. Below the critical
temperature a 3-brane with tension is formed, and this favors gravity localization.
This is the regime where warped compactification is established. Thus, if our universe
originates at very high temperatures, and if it appears to be five-dimensional, it may
become spontaneously compactified down to four dimensions below a critical temperature.
In this regime, supersymmetry can play the fundamental role of imposing that only
$AdS_4$ branes can be formed, instead of $M_4$ or $dS_4$ brane.

After reaching the critical temperature $T_c$ for brane formation, as the temperature
diminishes, the tension of the brane increases until a new phase transition occurs.
This is because of the fine-tuning between the brane tension $\sigma$ and the 5d
cosmological constant $\Lambda_5=-3/L^2$ that imposes conditions on the 4d cosmological
constant ${\Lambda}$. For $\sigma\to L|\Lambda_5|$ our
universe tends to a $M_4$ brane with cosmological constant ${\Lambda}\to0$.   
In this regime, the tower of masses $m_i^2$ decreases faster than the
4d cosmological constant, according to distinct power-laws. As the 4d cosmological constant
vanishes, the quasi-zero mode becomes a zero mode trapped on the $M_4$ brane \cite{rs}.
This leads to a smooth massless limit, which implies that no van Dam-Veltman-Zakharov
discontinuity exists \cite{porrati,kogan,kogan2}. The zero mode here is just at threshold
of the continuum spectrum of Kaluza-Klein modes. As the temperature decreases even more,
the brane tension increases until $\sigma>L|\Lambda_5|,$ and our universe becomes a $dS_4$
brane with cosmological constant ${\Lambda}>0,$ with the four-dimensional graviton
being a zero mode separated from the continuum spectrum. As we mentioned earlier,
a small positive 4d cosmological constant agrees with current observational data which
show that our universe experiences an accelerated expansion \cite{ac,ac2}. Thus, the phase
transition where our universe changes its geometry from an $AdS_4$ brane (${\Lambda}<0$)
to a $dS_4$ brane (${\Lambda}>0$) occurs at the critical temperature $T_*$.
The maximum value of $\sigma$ occurs for $BT^2<<2a^2\lambda^2,$ where $\Lambda$
stabilizes at some small positive constant.

The geometric phase transition which occurs at $T_*$ connects a scenario where supersymmetry
is present ($AdS_4$ brane) to another one, without supersymmetry ($dS_4$ brane). Thus,
it also indicates the presence of a soft supersymmetry breaking mechanism, which ``uplifts''
the 4d supersymmetric vacuum ${\Lambda}<0$ to a non-supersymmetric vacuum ${\Lambda}>0.$

The investigations that we have done are valid for a thick 3-brane. It can be extended
to two or more 3-branes where issues such as modulus stabilization should be
considered \cite{gw} --- see also \cite{odintsov} for high temperature effects
and modulus stabilization. Another issue is related to the recent investigation,
where two $AdS_4$ 3-branes with two gravitons which can vary their masses with the
position of the branes \cite{Shiyamala}. Other investigations are also of interest,
e.g., branes with internal structure \cite{bgomes1,bgomes2}, critical phenomena in
braneworlds \cite{campos} and gravity localization in scenarios with tachyon
potentials and supergravity braneworlds \cite{bbn2003} can be reconsidered
for $AdS_4$ and $dS_4$ branes.

\acknowledgments

We would like to thank PROCAD/CAPES and PRONEX/CNPq/FAPESQ for financial support.
DB thanks CNPq for partial support, FAB thanks Departamento de F\'\i sica, UFPB,
for hospitality, and ARG thanks FAPEMA for a fellowship.

\end{document}